\documentclass[doublecol]{epl2} 
\usepackage{graphicx}% Include figure files
\usepackage{amsmath}
\newcommand{\e}{\mathrm{e}}

\newcommand{\M}{\mathrm{M}}
\newcommand{\de}{\mathrm{d}}
\newcommand\Rey{\mbox{\textit{Re}}}  % Reynolds number

\title{Effective Medium Theory for Drag Reducing Micro-patterned Surfaces in Turbulent Flows}
\shorttitle{Effective Medium Theory for Drag Reducing Surfaces in Turbulent Flows} %Insert here a short version of the title if it exceeds 70 characters

\author{Ilenia Battiato\inst{1}}
\shortauthor{I. Battiato}

\institute{                    
  \inst{1}Mechanical Engineering Department, Clemson University - Clemson, SC, 29634
}

\abstract{
Inspired by the \textit{lotus effect}, many studies in the last decade have focused on micro-and nano-patterned surfaces. They revealed that patterns at the micro-scale combined with high contact angles can significantly reduce skin drag. However, the mechanisms and parameters that control drag reduction, e.g. Reynolds number and pattern geometry, are still unclear. We propose an effective medium representation of  the micro-features, that treats the latter as  a porous medium, and  provides a  framework to model flow over patterned surfaces in both Cassie and Wenzel states. Our key result is a closed-form expression for the skin friction coefficient in terms of frictional Reynolds  (or  K\'{a}rm\'{a}n) number in turbulent regime,  the viscosity ratio between the fluid in and above the features, and their geometrical properties.
We apply the proposed model to turbulent flows over superhydrophobic ridged surfaces. The  model predictions agree with laboratory experiments for  Reynolds numbers ranging from 3000 to 10000.}

\begin{document}

\maketitle

\section{Introduction}
The \textit{lotus effect} has  recently entered the scientific jargon to refer to superhydrophobicity. One of the necessary ingredients for a surface to be superhydrophobic is the presence of nano-  and micro-scale features. 
Patterned surfaces have shown drag-reducing abilities in both laminar \cite{cottin-2003-low,Joseph-2006-Slippage,choi-2006-effective,ybert-2007-achieving,lee-2008-structured} and turbulent \cite{sirovich-1997-turbulent,2009-daniello-drag} regimes, and in either Cassie  \cite{sirovich-1997-turbulent} or  Wenzel (or fakir)  state \cite{Joseph-2006-Slippage,choi-2006-effective,ybert-2007-achieving}. The former is characterised by the fluid impregnating the textured surface,  while in the latter the liquid interface  is suspended on an air cushion above the roughness peaks.

Drag reduction in turbulent flows by means of micropatterned superhydrophobic surfaces is of crucial importance because of its profound effects on a number of existing technologies. Yet, the optimal design of nano- and micro-scale roughness/structures for turbulent drag reduction is hampered by the relative lack of quantitative understanding of their impact on macroscopic flow observables, such as skin friction coefficient and  slip length. Also, the impact of Reynolds number is  unclear, and the upper limit of superhydrophobic drag reduction  is still unknown \cite{rothstein-2010-Slip}. Attempts to relate geometrical properties of the micro-features to macroscopic quantities, e.g. slip length, are mainly phenomenological  \cite{ybert-2007-achieving}, and analytical expressions are available only for tractable geometries \cite{Lauga-2003-effective,davis-2010-hydrodynamic}. 
Thus, a suitable  framework able to quantify effective properties of such surfaces and their connection to microscopic features  is still needed \cite{boucquet-2011-smooth}.

In this work we propose an effective-medium representation of Cassie- and Wenzel-like textured surfaces  in turbulent flows. We develop closed-form  expressions for the skin friction coefficient in terms of the geometrical properties of the patter and the friction Reynolds  (or  K\'{a}rm\'{a}n) number.  This is accomplished by treating the features as a porous medium.  The Reynolds equation for fully  turbulent channel flow over the pattern is coupled  to the porous media Brinkman equation for flow through the roughness.  We test the model veracity   by comparing  our closed-form expressions with skin friction data of turbulent channel flows over superhydrophobic grooved surfaces for Reynolds number ranging from 3000 to 10000  \cite{2009-daniello-drag}.

%%%%%%%%%%%%%%%%%%%%%%%%%%%%%%%%%%%%%
\section{Model Formulation}
%%%%%%%%%%%%%%%%%%%%%%%%%%%%%%%%%%%%%

We consider pressure-driven  channel flow through, $\hat y\in(-H,0)$, and over, $\hat y\in(0,2L)$, an array of   micro-ridges.  Following \cite{battiato-2010-Elastic,2012-battiato-selfsimilarity}, we treat the micro-patterned surface  as a porous medium with permeability $K$,   and couple Brinkman and  Reynolds equations to describe a distribution of the horizontal component of the average velocity $\hat u(\hat y)$ through, $\hat y\in(-H,0)$, and above the pattern, $\hat y\in(0,2L)$:
\begin{subequations}\label{eq:dimensional}
\begin{align}
&\mu_e \de_{\hat y\hat y} \hat u
 -\mu_e K^{-1}\hat u-\de_{\hat x}\hat p=0,& \hat y\in (-H,0)
\label{eq:brinkman_dimensional}\\
& \mu \de_{\hat y\hat y} \hat u
 -  \de_{\hat y} \langle \hat u'\hat v'\rangle -\de_{\hat x}\hat p=0,& \hat y\in
 (0,2L)\label{eq:channel_dimensional}
\end{align}
\end{subequations}
where  $\mu_e$ and $\mu$ are the dynamic viscosities of the fluids inside and above the porous medium (i.e. grooved surface), respectively.
 In the turbulent regime, $\de_{\hat x}\hat p<0$ is an externally imposed mean pressure gradient, $\hat{\mathbf u}=[\hat u, \hat v]$ denotes the mean velocity,  $\hat u'$ and $\hat v'$ are the velocity fluctuations about their respective means, and $\langle \hat u' \hat v'\rangle$ is the Reynolds stress. Fully-developed turbulent channel flow has velocity statistics that depend on $\hat y$ only. 
No-slip  is imposed at $\hat y=-H$ and $\hat y=2L$.  The formulation of appropriate boundary conditions at the interface between free and filtration (porous media) flows is still subject to open debate, which stems from the dispute of whether tangential velocity and shear stress at the interface are continuous or discontinuous~\cite{ochoa-1995-momentum,cieszko-1999-derivation,Jager-2000-interface,Weinbaum-2003-Mechanotransduction,lebars-2006-interfacial}. Following \cite{Weinbaum-2003-Mechanotransduction,battiato-2010-Elastic,2012-battiato-selfsimilarity} and many others, we impose the continuity of both velocity and shear stress at the interface, $\hat y=0$. Such conditions  have proven to provide accurate description of the macroscopic response of systems at the nanoscale \cite{Weinbaum-2003-Mechanotransduction,battiato-2010-Elastic,2012-battiato-selfsimilarity}. Hence, eq.~\eqref{eq:dimensional} are subject to 
\begin{align}
\hat u(-H) =0, & \quad \hat u(2L) =0, \,  \nonumber \\
\hat u(0^-)= \hat u (0^+) = \hat U,  & \quad 
\mu_e \de_{\hat y} \hat u |_{0^-}=  \mu \de_{\hat y} \hat u |_{0^+},\label{bc:continuity-shear_dimensionless}
\end{align}
where $\hat U$ is an unknown (slip) velocity at the interface $\hat y=0$.  
%The formulation of appropriate boundary conditions at the interface between free and filtration flows has been an active area of research since the seminal works by  \cite{beavers-1967-boundary} and \cite{taylor-1971-model}. 

Choosing $(L,\mu,q)$ as repeating variables, with $q=-L^2\de_{ \hat x}\hat p/ \mu$ a characteristic velocity, eq.~\eqref{eq:dimensional} can be cast in dimensionless form
\begin{subequations}\label{eq:dimensionless}
\begin{align}
& \M\de_{yy} u
 - \M\lambda^2 u+ 1=0,& y\in (-\delta,0)
\label{eq:brinkman_dimensionless}\\
& \de_{yy} u
 -  \Rey_\tau^2 \, \de_y \langle u'v'\rangle + 1=0,& y\in
 (0,2)%\label{eq:channel_dimensionless}
\end{align}
\end{subequations}
subject to $u(-\delta)=0$, $u(2)=0$, $u(0^-)=u(0^+)=U$, and $\M\de_y u |_{0^-}=\de_y u|_{0^+}$, where $y = \hat y/L$, $\delta=H/L$,  $\M=\mu_e/\mu$,   $u =\hat u/q$, and $U=\hat U/q$. The parameter $\lambda^2=(\M K)^{-1}L^2$ is inversely proportional to dimensionless permeability, $K/L^2$.  The limit $\lambda\rightarrow \infty$ corresponds to the diminishing flow through the  patterns due to decreasing permeability $K$. Furthermore, the K\'{a}rm\'{a}n (or frictional Reynolds) number, $\Rey_\tau$, is defined as the Reynolds number based on the channel half-width and the skin-friction velocity $\hat u_\tau=(-L\de_{\hat x}\hat p/\rho)^{1/2}$:
\begin{align}\label{eq:Re_tau}
\Rey_\tau:=\hat u_\tau L/\nu, \quad \mbox{or equivalently} \quad \Rey_\tau=(qL/\nu)^{1/2},
\end{align}
and it determines the relative importance of viscous and turbulent processes.  
Assuming the surface of the porous medium is hydrodynamically smooth, the law of the wall imposes $ \de_y u |_{0^+}=1$ since $u(y\rightarrow 0^+)=y+U$ in the viscous sublayer \cite{2012-battiato-selfsimilarity}. Therefore, inside the porous medium, i.e. $y\in[-\delta,0]$, the solution for the dimensionless velocity distribution $u(y)$ is given by  
\begin{subequations}\label{eq:brink-vel}
\begin{align}
& u(y) = (\M\lambda^{2})^{-1} + C_1 \e^{\lambda y} + C_2 \e^{-\lambda y}, \label{eq:brink-velocity-only}\\
& C_{1,2} =\pm\dfrac{1}{\M\lambda^2}\dfrac{(\M\lambda^2U-1)\e^{\pm\delta\lambda}+1}{
 \e^{\delta\lambda}-\e^{-\delta\lambda}},\label{brinkman:constants} \\
 & U = (\M\lambda^{2})^{-1} (1+\lambda\tanh{\delta\lambda}-\mbox{sech}\,\delta \lambda).\label{brinkman:U}
 \end{align}
\end{subequations}

The skin friction coefficient is defined as $C_f=2\hat \tau_0/(\rho \hat{u}_{\mathrm b}^2)$,
where $\hat \tau_0=\mu\de_{\hat y} \hat u|_{0^+}$ is the shear stress at the edge of the pattern,  $\hat u_{\mathrm b} =q \chi$ is the average flow velocity, and $\chi=(2+\delta)^{-1}\int_{-\delta}^2u(y)\de y$ is a dimensionless bulk velocity. From \eqref{eq:Re_tau}, 
\begin{align} \label{eq:q}
q=\dfrac{\nu\Rey_\tau^2}{L}.
\end{align}
Then, $\hat u_{\mathrm b}=\nu \Rey_\tau^2 \chi /L$ and the skin friction coefficient is written in terms of $\Rey_\tau$, 
\begin{align}\label{eq:C_f-general-new}
C_f(\Rey_\tau)=\dfrac{2}{\Rey_\tau^2\chi^2}
\end{align}
since $\de_y u|_{0^+}=1$. The dimensionless bulk velocity  $\chi$ is rearranged as follows,
\begin{align}\label{eq:chi}
\chi= \dfrac{\chi_\delta+ 2\chi_t}{2+\delta}, \quad \chi_\delta=\int_{-\delta}^0 u(y)\de y, \quad \chi_t=\frac{1}{2}\int_0^2 u(y)\de y.
\end{align}
%
%where $\chi_\delta=\int_{-\delta}^0 u(y)\de y$ and $\chi_t=\frac{1}{2}\int_0^2 u(y)\de y$. 
Equation~\eqref{eq:chi} shows the impact  of the pattern on the skin friction coefficient: $\chi\equiv\chi_t$  when $\delta=0$, i.e. for a smooth  channel. 
Integrating  \eqref{eq:brink-velocity-only},  and combining the result with \eqref{brinkman:constants} and \eqref{brinkman:U}, we obtain
\begin{align}
\chi_\delta=(\M\lambda^{3})^{-1}\left[\lambda(1+\delta)+ \mbox{sech}\Lambda\left(\mbox{csch} \Lambda-\lambda\right) - \mbox{coth}\Lambda\right],
\end{align}
with $\Lambda=\lambda\delta$. The scale parameter $\Lambda$ provides a formal classification between thin ($\Lambda\ll1$) and thick ($\Lambda\gg1$) porous media \cite{2012-battiato-selfsimilarity}. Since the pattern vertical length scale is generally very small compared to the height of the channel, i.e. $\delta\rightarrow 0$, we look for the asymptotic behaviour of $\chi_\delta$ as $\Lambda\rightarrow 0$. In this limit,
\begin{align}\label{eq:asympt-u-delta-turbulent}
\chi_\delta \sim \dfrac{\delta}{\M\lambda^2}.
\end{align}
Assuming that the effect of the slip velocity $\hat U$ on the bulk velocity in the channel $q\chi_t$ is negligibly small when $\delta\rightarrow 0$ \cite[Fig.1(a)]{fukagata-2006-theoretical}, we employ the \emph{log-law} and the velocity-defect law of turbulent flow in a channel of width $2L$ to provide an estimate for $\chi_t$. These two laws combined relate the friction velocity $\hat u_\tau$ to the channel bulk velocity $q\chi_t$,  % \cite[][Eqs.~(7.52) and~(7.54)]{Pope},
\begin{align}\label{eq:defect-law}
\dfrac{1}{\kappa}+\dfrac{q\chi_t}{\hat u_\tau}=\dfrac{\ln Re_\tau}{\kappa}+5.1,
\end{align}
where $\kappa=0.41$ is the von K\'{a}rm\'{a}n constant.
Inserting~\eqref{eq:q} in~\eqref{eq:defect-law}, we obtain
\begin{align}\label{eq:average-turbulent2}
\chi_t(\Rey_\tau)=\dfrac{\ln \Rey_\tau +5.1\kappa-1}{\kappa \Rey_\tau},%\left(\ln \Rey_\tau +5.1\kappa-1\right).
\end{align} 
since $\hat u_\tau =\nu Re_\tau/L$.
%
%since $\mbox{sech}\Lambda\sim 1$,  $\mbox{csch}\Lambda\sim \Lambda^{-1}$, and $\mbox{coth}\Lambda\sim \Lambda^{-1}$ as $\Lambda\rightarrow 0$. 
Combining~\eqref{eq:C_f-general-new}, \eqref{eq:chi}, \eqref{eq:asympt-u-delta-turbulent} and  \eqref{eq:average-turbulent2}   we obtain a closed form expression for the skin friction coefficient in terms of  the viscosity ratio between the fluids inside and over the patterns, $\M$,  K\'{a}rm\'{a}n number, $\Rey_\tau$, and the pattern height and effective permeability, $\delta$ and $\lambda$, respectively,
\begin{subequations}\label{Cf_tot}
\begin{align}
C_f=C_{f}^s\left(\dfrac{2+\delta}{2+\mathcal T \delta}\right)^2 \label{C_f-turbulent} 
\end{align}
with
\begin{align}
\mathcal T=\dfrac{1}{\M\lambda^2 \chi_t(\Rey_\tau)}, \quad C_{f}^s=2\left(\dfrac{\kappa}{\ln\Rey_\tau+5.1\kappa-1}\right)^2.\label{C_f-smooth}
\end{align}
%and
%\begin{align}
%C_{f}^s=2\left(\dfrac{\kappa}{\ln\Rey_\tau+5.1\kappa-1}\right)^2.\label{C_f-smooth}
%\end{align}
\end{subequations}
Here $C_f^s$ represents the skin friction coefficient in a channel with smooth walls. 
Similarly, for two patterned walls, the skin friction coefficient, $C_{f2}$, is
\begin{align}\label{C_f-turbulent2walls}
C_{f2}=C_{f}^s\left(\dfrac{1+\delta}{1+\mathcal T \delta}\right)^2.
\end{align} 

The turbulent drag reduction $R_D^\%=(1-C_f/C_f^s)\%$ for a channel with one (or two) superhydrophobic walls can be readily calculated from  \eqref{C_f-turbulent} (or \eqref{C_f-turbulent2walls}):
\begin{align}\label{R_D1wall}
R_D^\%=100-\left(\dfrac{2+\delta}{2+\mathcal T \delta}\right)^2\%.
\end{align} 
Equations~\eqref{Cf_tot} and~\eqref{C_f-turbulent2walls}  provide closed-form expressions for  $C_f$ whenever the effective permeability of the micro pattern, the geometry of the channel and the operational flow conditions are known. 

%%%%%%%%%%%%%%%%%%%%%%%%%%
\section{Comparison with experiments}
%%%%%%%%%%%%%%%%%%%%%%%%%%
We test the robustness of our  model by comparing it  with experiments \cite{2009-daniello-drag}. Data sets collected by Ref. \cite[Figs.~8 and 9]{2009-daniello-drag} include  measurements of skin friction and drag reduction coefficients, $C_f$ and $R_D^\%$ respectively,  as a function of Reynolds number ($\Rey=2L\hat u_{\mathrm b}/\nu$) from channels with  smooth walls, and one  and two superhydrophobic walls containing 30$\mu$m wide microridges spaced 30$\mu$m apart. A set of dimensional and dimensionless parameters for the experiments are listed in table~\ref{table:samples}. The data  span almost one order-of-magnitude wide range of Reynolds number both in laminar and turbulent regimes,  $\Rey\in(2000,10000)$.

\begin{figure}%[htbp]
\onefigure[width=8.8cm]{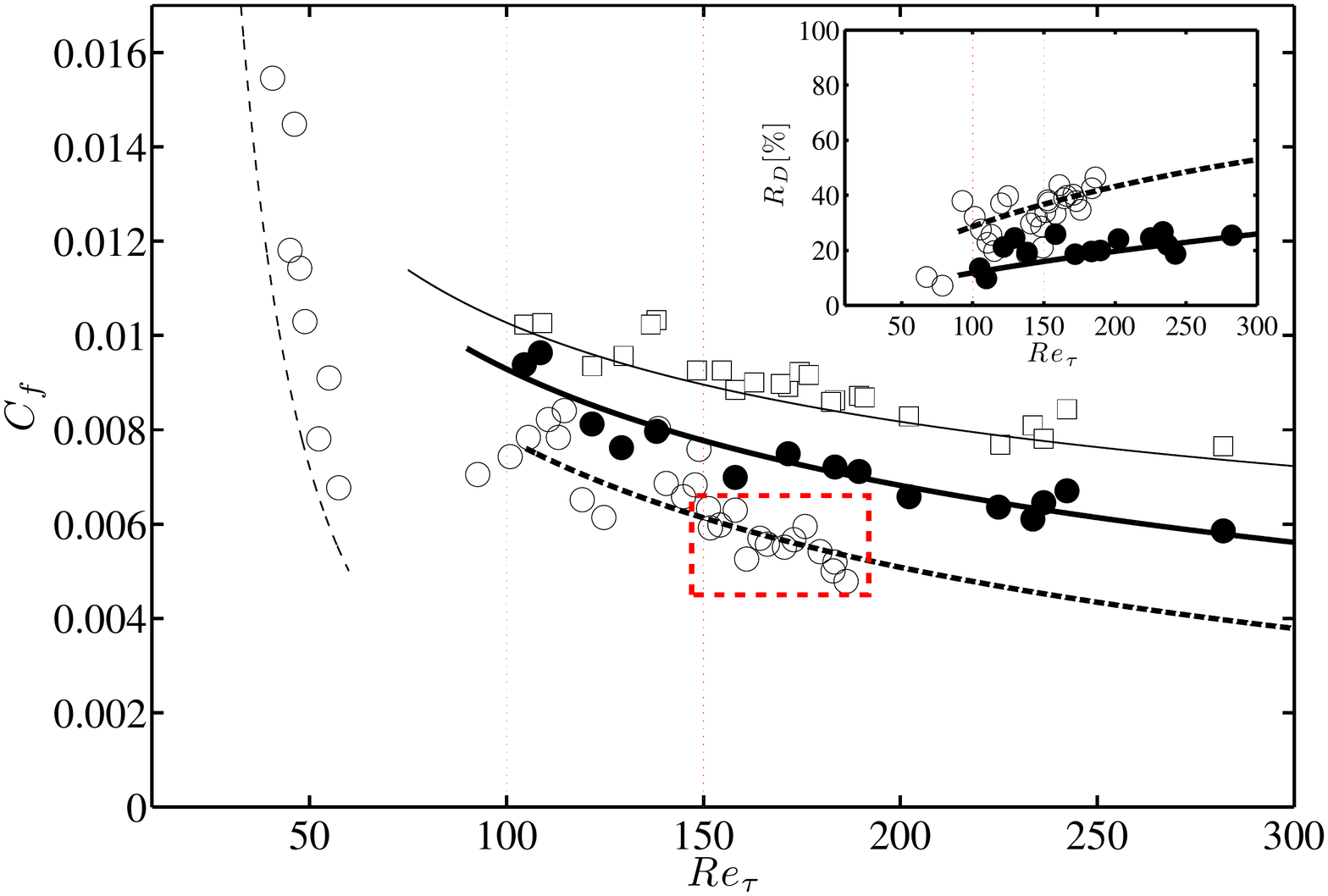}
 \caption{Experimental  (symbols) and predicted (lines) skin friction $C_f$ in terms of $\Rey_\tau$.  Data adapted from \cite[Fig.8]{2009-daniello-drag}. Measurements of skin friction coefficient for a channel with smooth  walls (empty squares), one  (filled circles)  and two  (empty circles) SHS with 30$\mu$m ridges spaced 30$\mu$m apart. The thin dashed and solid lines represent the theoretical prediction of the skin friction coefficient for smooth channel, $C_f^s$, in laminar and turbulent regimes given by $C^s_f=18/\Rey_\tau^2$ and~\eqref{C_f-smooth},  respectively. The thick dashed and solid lines represent a one-parameter fit ($\lambda=4.54$) and a parameter-free prediction of $C_f$ given by  \eqref{C_f-turbulent2walls} and \eqref{C_f-turbulent}, respectively. The dashed box contains the data used for the parametric fitting. Inset: Experimental (symbols) and predicted (lines) drag reduction in terms of $\Rey_\tau$.  Data adapted from \cite[Fig.9]{2009-daniello-drag}.
 }
 \label{fig:experiment}
\end{figure}

A  comparison between model and experiments requires one to establish a relationship between $\Rey$ and $\Rey_\tau$. Combining \eqref{eq:q} with the definition of dimensionless bulk velocity $\chi$ (i.e. $q=\hat u_{\mathrm b}\chi^{-1}$), and multiplying both sides by $2L/\nu$ leads to
\begin{align}\label{eq:Re-Re_tau}
\Rey=2\chi\Rey_\tau^2.
\end{align}
For turbulent smooth-channel  flows, combining  \eqref{eq:average-turbulent2} with \eqref{eq:Re-Re_tau} leads to a relation between $\Rey$ and $\Rey_\tau$ in the form $\Rey=2\Rey_\tau(\kappa^{-1}\ln \Rey_\tau+5.1-\kappa^{-1})$. Similarly, for a channel with one (or two) micro-patterned surfaces, inserting \eqref{eq:chi} into  \eqref{eq:Re-Re_tau}  leads to $\Rey=2\Rey_\tau^2(2+\delta)^{-1}(\chi_\delta+2\chi_t)$ (or  $\Rey=2\Rey_\tau^2(1+\delta)^{-1}(\chi_\delta+\chi_t)$). Since $\chi_\delta\ll \chi_t$ as $\delta\rightarrow 0$, a good approximation of the former equations  is $\Rey_\tau\approx 0.09\Rey^{0.88}$ for $\Rey<4\cdot10^4$. Additionally, in laminar smooth-channel  flows the  dimensionless parabolic velocity profile, $u(y)=-y^2+y$, $y\in[0,2]$, gives $\chi=1/3$. When combined with (7) and (16), this leads to the well-known skin friction formula $C_f=12/\Rey$ or, equivalently, $C_f=18/\Rey_\tau^2$.
 The former relationships allow us to rescale the data points from  \cite{2009-daniello-drag} as showed in Fig.~\ref{fig:experiment}. Transitional effects from laminar to turbulent regimes are  apparent in the range $\Rey_\tau\in[100,150]$ for  channel flow with two superhydrophobic surfaces. 

Except for relatively simple configurations (e.g. an array of pillars \cite{battiato-2010-Elastic}), there exist no exact closed-form expressions that relate the dimensionless effective permeability $\lambda$ to the geometrical properties of riblets. Therefore, we validate the proposed model by employing two sets of independent measurements from \cite{2009-daniello-drag}. The first dataset consists of measurements of the skin friction coefficient $C_f$ in a channel with two micro-patterned walls, for the fully turbulent regime represented by a range of K\'{a}rm\'{a}n number $\Rey_\tau\in[150,200]$ (see the dash-lined box in fig.~\ref{fig:experiment}). Fitting to these data yields the value of permeability $\lambda = 4.54$. This value is used to make fit-free predictions of the skin friction coefficient $C_f$ in a channel with one smooth wall and one micro-patterned wall, for the fully turbulent regime represented by a range of K\'{a}rm\'{a}n number $\Rey_\tau\in[100,300]$. Figure~\ref{fig:experiment}, which compares this prediction (bold solid line) with the corresponding $C_f$ measurements (filled dots) comprising the second dataset, shows a good agreement between data and model solution.

The fitted value of $\lambda$ corresponds to the permeability $K=1.8\cdot 10^{-5}\mbox{m}^2$ of the effective porous medium used to represent the two 30$\mu$m-ridged superhydrophobic walls. An order-of-magnitude analysis of the permeability of this porous medium is obtained from Darcy's law, which states that the Darcy flux $\hat q_d$ (volumetric flow rate per unit height $H$) is proportional to $|\de_{\hat x}\hat p |$, the applied pressure gradient, such that
\begin{align}\label{eq:permeability}
K=\dfrac{\mu_e}{|\de_{\hat x}\hat p |}\hat q_d.
\end{align}

 Each patterned surface in the experimental setup~\cite{2009-daniello-drag} is $38.1$mm wide, consisting of an array of $n \approx 635$ square ridges of height $H = 30$ $\mu$m spaced 30 $\mu$m apart. %Similarly, channels with two superhydrophobic walls contain a total of 1270 30$\mu$m ridges.
We approximate the flow between any two ridges with a  fully-developed  pressure-driven flow between two parallel plates the distance $H$ apart; the bottom plate is fixed while the upper plate moves with a uniform speed $\hat U^\star=(1-\phi_s)^{-1}\hat U$ where $\hat U$ is the slip velocity measured in~\cite{2009-daniello-drag} and $\phi_s=0.5$ is the solid fraction of the patterned surface. Then $\hat q_d=\hat U^\star/2-H^2\de_{\hat x} \hat p/\mu_e$, and \eqref{eq:permeability} gives the permeability of an individual channel $K_i=-\mu_e \hat U^\star/(2 \de_{\hat x}\hat p)+H^2/12$ ($i =1,\ldots,n$).  The total permeability of the two patterned surfaces is $K = 2 \sum_{i=1}^n K_i$. 

In~\cite[Fig.5b]{2009-daniello-drag}, the slip velocity $\hat U=0.2$ ms$^{-1}$ is reported  for the channel with two patterned walls (square ridges of $H = 30$ $\mu$m) and $\Rey=7930$. In the absence of reported pressure measurements for this channel configuration, we employ the pressure drop data reported for two other channels~\cite[Fig.6]{2009-daniello-drag}. In the  first channel (two smooth walls) the pressure drop was $|\de_{\hat x}\hat p| = 2.6$ kPa$\cdot$m$^{-1}$. In the second (both surfaces patterned with $H = 60\mu$m square ridges) it was $|\de_{\hat x}\hat p| = 1.4$ kPa$\cdot$m$^{-1}$. Using these two values as upper and lower bounds for the actual $|\de_{\hat x}\hat p|$, we obtain permeability bounds $1.8\cdot10^{-6}$~m$^2$ $\le K \le 3.3\cdot 10^{-6}$~m$^2$. These estimates differ by a factor of $5-10$ from the fitted value of $K=1.8\cdot 10^{-5}\mbox{m}^2$. The discrepancy between the two is to be expected due to deviations of the experiment from the model approximations and/or highly idealised conditions, which include, e.g., flow steadiness and one-dimensionality, and  hydrodynamically smoothness of the ridges' tips.

%%%%%%%%%%%%%%%%%%%%%%%%

%
%\begin{table}
%\caption{Table caption.}
%\label{tab.1}
%\begin{center}
%\begin{tabular}{lcr}
%first  & table & row\\
%second & table & row
%\end{tabular}
%\end{center}
%\end{table}
%%%%%%%%%%%%%%%%%%%%%%%%
\begin{table}
\caption{Parameter values used in the experiments of \cite{2009-daniello-drag} with  channels with smooth walls, one and two superhydrophobic surfaces with 30$\mu$m ridges spaced 30$\mu$m apart. Dimensionless quantities are calculated from corresponding dimensional parameters.}\label{table:samples}
\begin{center}
%\vspace{0.5cm}
\begin{tabular}{lccr}%{lllll}
%\hline
Sample \qquad & Smooth \qquad & 1 SHS\qquad &   2 SHS    \qquad \\ \hline \\

$L$ [m] &   $3.95\cdot 10^{-3}$ &   $3.95\cdot 10^{-3}$   &      $2.75\cdot 10^{-3}$ \\
$H$ [m] &   $0$ &   $25\cdot 10^{-6}$   &      $25\cdot 10^{-6}$ \\
$\mu$ [Pa$\cdot$s] & $8.90\cdot 10^{-4}$ \qquad  &  $8.90\cdot 10^{-4}$ \qquad &  $8.90\cdot 10^{-4}$  \qquad    \\
$\mu_e$ [Pa$\cdot$s] \qquad\qquad &   $8.90\cdot 10^{-4}$ &   $1.78\cdot 10^{-5}$   &      $1.78\cdot 10^{-5}$ \\
$\delta$ [-] &   $0$ &   $6.33\cdot 10^{-3}$   &      $9.01\cdot 10^{-3}$ \\
$\M$ [-] &   $1$ &   $0.02$   &      $0.02$ \\
%$\lambda$ [-] & $\infty$ \qquad  &  $4.54$ \qquad &   $4.54$      \\
%\hline
\end{tabular}
\end{center}
\end{table}
%%%%%%%%%%%%%%%%%%%%%%%%%%

Next, we discuss some  implications of the former model.
Equation~\eqref{C_f-turbulent} implies that $C_f<C_f^s$ if $\mathcal T>1$,
or
\begin{align}\label{condition}
C_f<C_f^s \quad \mbox{if} \quad \chi_t^{-1}(\Rey_\tau)>\M\lambda^2,
\end{align}
with $\chi_t$ defined by \eqref{eq:average-turbulent2} and $\Rey_\tau>\Rey_\tau^{\mathrm t}$, with $\Rey_\tau^{\mathrm t}$ the transition K\'{a}rm\'{a}n number between laminar and turbulent regimes. For channel flow,  $\Rey^{\mathrm t}_{\tau}\approx 100$ (or $\Rey^{\mathrm t} \approx 3000$). At  any fixed K\'{a}rm\'{a}n number, the skin friction $C_f$ is smaller than its smooth channel counterpart when appropriate conditions of the roughness/pattern geometry, $\lambda$ and $\delta$, and of the fluids,  $\M$, are met. 
Also, since $\chi_t^{-1}(\Rey_\tau)$ is a convex function,  \eqref{condition} implies the following:\\

\textbf{Proposition.} For any fixed  configuration of obstacles, $\lambda$, and fluid viscosity ratio, $\M$, such that $\M\lambda^2\geq \chi_t^{-1}(\Rey^{\mathrm t}_\tau)\approx{7.2}$, there exists  a critical K\'{a}rm\'{a}n number, $\Rey_\tau^\star$, such that $C_f\leq~C_f^s$ if $\Rey_\tau>\Rey_\tau^\star$ where $\Rey_\tau^\star$ is a root of the transcendental equation
\begin{align}
\kappa \Rey^\star_\tau\left(\ln \Rey^\star_\tau +5.1\kappa-1\right)^{-1}=\M\lambda^2, \quad \Rey^\star_\tau>\Rey_\tau^{\mathrm t}.
\end{align}

The existence of a $\Rey^\star_\tau$ is consistent with experimental results, where drag reduction is initiated at a critical Reynolds number, just past the transition to turbulent flow \cite{2009-daniello-drag}.  The former statement can be reformulated as a condition on the geometrical properties of the patterns/roughness, $\lambda$, and the viscosity ratio, $\M$: for any fixed value of K\'{a}rm\'{a}n number $\Rey_\tau^0>\Rey_\tau^{\mathrm t}$, drag reduction is achieved if the product $\M\lambda^2$  is  bounded from below and above, i.e. 
\begin{align}\label{eq:permeability-ebounds}
\chi_t^{-1}(\Rey_\tau^{\mathrm t})<\M\lambda^2<\chi_t^{-1}(\Rey_\tau^0), \quad \Rey_\tau^0>\Rey_\tau^{\mathrm t}
\end{align}
with $\chi_t$ defined in \eqref{eq:average-turbulent2}, and $\chi_t^{-1}(\Rey_\tau^{\mathrm t})\approx 7.2$. 

This analysis has the following implications. (i) The proposed model suggests that drag reduction is achieved when $\lambda>1$, i.e. in the \emph{porous medium} regime \cite{2012-battiato-selfsimilarity}, and for an intermediate range of effective permeability values. The upper bound on $\lambda$ (i.e. the minimum value of permeability) is determined by the magnitude of $\Rey^0_\tau$, i.e. the operational flow conditions of the apparatus/system. This is consistent with passive turbulent flow control systems where porous surfaces in airfoils are employed for drag reduction purposes.  (ii) The transition between  drag enhancing and  reducing regimes is governed by the geometric parameters of the obstacles, $\lambda$, and the viscosity of the fluid flowing between the roughness/pattern and above it, $\M$. (iii) For any fixed geometry and  $\Rey_\tau>\Rey_\tau^\star$, lower drag is achieved in Cassie/Fakir state  than in Wenzel state  since $\M<1$ in the former case. Also this result is consistent with experimental observations. While the former observations are qualitatively consistent with experiments, future work will focus on a quantitative analysis/estimate of each process  above mentioned. 

%%%%%%%%%%%%%%%%%%%%%%%%%
\section{Concluding remarks}
%%%%%%%%%%%%%%%%%%%%%%%%%

We proposed a novel continuum-scale framework to modelling turbulent flows over micro-patterned surfaces. While applicable to  flows over patterned surfaces both in Cassie and Wenzel state, we test the model   on turbulent flows over superhydrophobic ridged surfaces. %Our model goes beyond phenomenological approaches, which fit data from physical experiments \cite{ybert-2007-achieving,cheng-2009-microchannel}.  
To the best of our knowledge, this is the first continuum-scale framework that allows one to successfully  quantify and analytically predict  the impact of pattern geometry and Reynolds number on drag reduction. This is achieved by modelling the micro-patterned surface as a porous medium,  and by coupling Brinkman equation for flow in porous media with Reynolds equations, which describe the average flow  through and over the pattern, respectively. This yields a closed-form solution for the skin friction coefficient in terms of the frictional Reynolds (K\'{a}rm\'{a}n) number, the viscosity ratio between the outer and inner fluid, and the geometrical (i.e. height) and effective properties (i.e. permeability) of the micro-structure. We demonstrated good agreement between our model and experimental data.

%We analytically derive dynamical (i.e. critical K\'{a}rm\'{a}n number) and geometrical/structural (i.e. viscosity ratio, pattern's height and permeability) conditions under which drag reduction can be achieved. 
%Finally, the theoretical model suggests that  drag reduction is achieved in the porous medium regime, and for an intermediate range of effective permeability values. 
Based on dynamical and geometrical conditions under which the proposed model predicts drag reduction, we conjecture that the latter might be attributed to a \emph{porous-like} medium behaviour of the roughness/pattern. We speculate that our results might provide an insight on the transition between turbulent flows over drag-increasing  \cite{castro-2007-rough}  and drag-decreasing rough walls where patterned protrusions, rigid or compliant \cite{sirovich-1997-turbulent,Bruecher-2011-interaction} to the flow, or porous coatings \cite{2012-venkataraman-numerical}, can be used to attenuate near wall turbulence. Yet, the connection between the flow characteristics at the pattern-scale and their effective-medium behaviour needs to be elucidated and is subject of current investigations. 
%====================
%Insert here the text.
%See fig.~\ref{fig.1}, table~\ref{tab.1} and eq.~(\ref{eq.1}).
%See also~\cite{b.a,b.b}.
%\begin{equation}
%\label{eq.1}
%0\neq1
%\end{equation}
%
%

\acknowledgments
Part of this research was developed when the author was first a postdoctoral fellow at Max Planck Institute for Dynamics and Self-Organization (MPI-DS), G\"{o}ttingen, 37077, Germany, and later a visiting scientist at the Statistical and Applied Mathematical Sciences Institute (SAMSI), Research Triangle Park, NC 27709, USA.

  \bibliographystyle{eplbib}
\bibliography{jfm2esam}

\end{document}